\documentstyle[nato,numreferences]{crckapb}

\begin{opening}
\title{PECULIARITIES OF THE SPECTRUM\protect\\
       OF STRONGLY CORRELATED ELECTRONS}

\author{A.~Sherman}
\institute{Institute of Physics, University of Tartu, Riia 142,\\
           51014 Tartu, Estonia}
\author{M.~Schreiber}
\institute{Institut f\"ur Physik, Technische Universit\"at,\\
           D-09107 Chemnitz, Federal Republic of Germany\\
           and\\
           School of Engineering and Science,\\
           International University Bremen, Campus Ring 1,\\
           D-28759 Bremen, Federal Republic of Germany}
\end{opening}

\begin{document}

\begin{abstract}
Hole and spin Green's functions of the two-dimensional $t$-$J$ model of
the CuO$_2$ planes are calculated in an approximation which retains the
rotation symmetry in the paramagnetic state and has no presumed magnetic
ordering. In this approximation Green's functions are represented by
continued fractions which are interrupted by decouplings with a vertex
correction determined from the constraint of zero site magnetization in
the paramagnetic state. Obtained results are shown to be in good
agreement with Monte Carlo and exact diagonalization data. The
calculated spectra demonstrate a pseudogap and a band splitting which
are similar to those observed in Bi-based cuprate perovskites.
\end{abstract}

For the description of strongly correlated electrons in CuO$_2$ planes
of perovskite high-$T_c$ superconductors the two-dimensional $t$-$J$
model is widely used (for a review see Ref.~\cite{izyumov}). One of the
most fruitful analytical methods for this model uses the spin-wave
description of magnetic excitations. Some spectral peculiarities
obtained in this approach are similar to those revealed from the
photoemission, spin-lattice relaxation and neutron scattering
experiments (see, e.g., Ref.~\cite{sherman98}). However, this
approximation has two serious shortcomings: (i) deviations from the
N\'eel state are presumed to be small and (ii) the approximation
violates the rotation symmetry of the paramagnetic state.

In this paper we try to overcome these difficulties by using the
description in terms of spin and hole operators and the continued
fraction representations for Green's functions. Self-energies are
calculated by decouplings corrected with the use of the constraint of
zero site magnetization in the paramagnetic state. The Hamiltonian of
the $t$-$J$ model reads
\begin{equation}
H=\sum_{\bf nm\sigma}t_{\bf nm}a^\dagger_{\bf n\sigma} 
a_{\bf m\sigma}+\frac{1}{2}\sum_{\bf nm}J_{\bf nm}\left( 
s^z_{\bf n}s^z_{\bf m}+s^{+1}_{\bf n}s^{-1}_{\bf m}\right)+\mu 
\sum_{\bf n}X_{\bf n},
\label{hamiltonian}\end{equation}
where $a_{\bf n\sigma}=|{\bf n}\sigma\rangle\langle{\bf n}0|$ is the
hole destruction operator, {\bf n} and {\bf m} label sites of the square
lattice, $\sigma=\pm 1$ is the spin, $|{\bf n}\sigma\rangle$ and $|{\bf
n}0\rangle$ are site states corresponding to the absence and presence of
a hole on the site. In these notations the spin-$\frac{1}{2}$ operators
are written as $s^z_{\bf
n}=\frac{1}{2}\sum_\sigma\sigma|{\bf n}\sigma\rangle\langle{\bf
n}\sigma|$ and $s^\sigma_{\bf n}=|{\bf n}\sigma\rangle\langle{\bf
n},-\sigma|$. The hopping and exchange constants $t_{\bf nm}$ and
$J_{\bf nm}$ are expected to be nonzero only for nearest neighbor sites
for which they take the values $t$ and $J$. $\mu$ is the chemical
potential and $X_{\bf n}=|{\bf n}0\rangle\langle{\bf n}0|$.

The hole and spin Green's functions are determined as follows:
\begin{equation}
G({\bf k}t)=-i\theta(t)\langle\!\{a_{\bf k\sigma}(t),
 a^\dagger_{\bf k\sigma}\}\!\rangle,\quad
D({\bf k}t)=-i\theta(t)\langle[s^z_{\bf  k}(t),s^z_{\bf -k}]\rangle,
\label{green}\end{equation}
where angular brackets denote canonical ensemble averaging, $a_{\bf
k\sigma}=N^{-1/2}$ $\sum_{\bf n}\exp(-i{\bf kn})a_{\bf n\sigma}$ and
$s^z_{\bf k}=N^{-1/2}\sum_{\bf n}\exp(-i{\bf kn})s^z_{\bf n}$ with the
number of sites $N$. To calculate Green's functions (\ref{green}) we use
their continued fraction representation \cite{mori,sherman87},
\begin{equation}
\langle\!\langle A_0|A_0^\dagger\rangle\!
 \rangle_\omega=\frac{\displaystyle|
 A_0\cdot A_0^\dagger|}{\displaystyle\omega-E_0-
 \frac{\displaystyle F_0}{\displaystyle\omega-E_1-
 \frac{\displaystyle F_1}{\ddots}}},
\label{cfraction}\end{equation}
where $\langle\!\langle A_0|A_0^\dagger\rangle\!\rangle_\omega$ is the
Fourier transformation of one of the above Green's functions and the
operator $A_0$ is equal to $a_{\bf k\sigma}$ or $s^z_{\bf k}$. The
coefficients $E_i$ and $F_i$ of continued fraction (\ref{cfraction}) are
determined by the recursive procedure
\begin{eqnarray}
&&[A_n,H]=E_nA_n+A_{n+1}+F_{n-1}A_{n-1},\nonumber\\
&&\label{lanczos}\\
&&E_n=|[A_n,H]\cdot A_n^\dagger|\,|A_n\cdot A_n^\dagger|^{-1},\;
 F_{n-1}=|A_n\cdot A_n^\dagger|\,|A_{n-1}\cdot A_{n-1}^\dagger|^{-1}.
 \nonumber
\end{eqnarray}
The definition of the inner product $|A\cdot B^\dagger|$ depends on the
considered function. In the case of the anticommutator Green's function
$G({\bf k}\omega)$ in Eq.~(\ref{green}) this inner product is defined as
$\langle\!\{A,B^\dagger\}\! \rangle$, while for the commutator function
$D({\bf k}\omega)$ it is $\langle[A,B^\dagger]\rangle$. However, in this
latter case for the particular operator $A_0=s^z_{\bf k}$ the numerator
of the continued fraction in Eq.~(\ref{cfraction}) vanishes. Therefore
the direct application of Eqs.~(\ref{cfraction}) and (\ref{lanczos}) for
the calculation of $D({\bf k}\omega)$ is impossible. Instead we consider
Kubo's relaxation function
\begin{equation}
(\!(s^z_{\bf k}|s^z_{\bf -k})\!)_t=\theta(t)
 \int^\infty_t dt'\langle[s^z_{\bf k}(t'),s^z_{\bf -k}]\rangle.
\label{kubo}\end{equation}
Continued fraction representation (\ref{cfraction}) with recursive
procedure (\ref{lanczos}) is also valid for this relaxation function, if
the inner product is defined as $(A,B^\dagger)=i \int_0^\infty dt\,
\exp(-\eta t)\langle[A(t),B^\dagger] \rangle$, $\eta\rightarrow +0$.
On calculating the relaxation function, Green's function is found from
the relation
\begin{equation}
\omega(\!(s^z_{\bf k}|s^z_{\bf -k})\!)=\langle\!\langle s^z_{\bf k}|
 s^z_{\bf -k}\rangle\!\rangle+(s^z_{\bf k},s^z_{\bf -k}).
\label{kg}\end{equation}

For reasons of space we shall not go into details of the calculations
and give only final results (a more detailed discussion will be
published elsewhere). The spin Green's function reads
\begin{equation}
D({\bf k}\omega)=\frac{2\omega\,(s^z_{\bf k},s^z_{\bf -k})\,
 \Pi({\bf k}\omega)+4JC_1(\gamma_{\bf k}-1)}{\omega^2
 -2\omega\,\Pi({\bf k}\omega)-\omega^2_{\bf k}},
\label{sgf}\end{equation}
with the square of the frequency of magnetic excitations
\begin{equation}
\omega^2_{\bf k}=16J^2\alpha|C_1|(1-\gamma_{\bf k})(\Delta+1+
 \gamma_{\bf k})
\label{freq}\end{equation}
and the imaginary part of the polarization operator
\begin{eqnarray}
&&{\rm Im}\,\Pi({\bf k}\omega)=\frac{\pi}{\omega}\sum_{\bf k'}f^2_{\bf
 k'k}\int^\infty_{-\infty}d\omega'[n_F(\omega')-n_F(\omega'-\omega)]
 \nonumber\\
&&\qquad\qquad\;\;\times A({\bf k'-k},\omega'-\omega)A({\bf k'}\omega')
\label{po}\end{eqnarray}
(the real part of $\Pi({\bf k}\omega)$ is determined from the dispersion
relations). In the above equations, $(s^z_{\bf k},s^z_{\bf
-k})^{-1}=4J\alpha(\Delta +1+\gamma_{\bf k})$,
$\Delta=C_2/|C_1|+(1-\alpha)/(8\alpha|C_1|)-3/4$, $\gamma_{\bf
k}=[\cos(k_x)+\cos(k_y)]/2$, $n_F(\omega)=[\exp(\omega/T )+1]^{-1}$, $T$
is the temperature, $f_{\bf k'k}= 2tN^{-1/2}(\gamma_{\bf k'}-\gamma_{\bf
k'-k}) (s^z_{\bf k},s^z_{\bf -k})^{-1/2}$, and the hole spectral
function $A({\bf k}\omega)=-\pi{\rm Im}\,G({\bf k}\omega)$. The
parameters $C_p= N^{-1}\sum_{\bf k}\gamma^p_{\bf k}C_{\bf k}$ with
$p=1,2$ and $C_{\bf k} =\sum_{\bf n}\exp[i{\bf k(n-m)}]\langle
s^{+1}_{\bf n}s^{-1}_{\bf m} \rangle$ appear in the above formulas after
the decoupling of correlations containing four spin operators. These
correlations arise in the second step of recursive procedure
(\ref{lanczos}) and the decoupling serves as the terminator for
continued fraction (\ref{cfraction}). Following Ref.~\cite{kondo} we
somewhat improve this approximation by multiplying the result of the
decoupling with the vertex correction $\alpha$. The parameters $C_1$,
$C_2$ and $\alpha$ are calculated from the above definition of $C_p$ and
the constraint of zero site magnetization in the paramagnetic state,
\begin{equation}
\langle s^z_{\bf n}\rangle =\frac{1}{2}\left(1-x\right)-\langle 
s^{-1}_{\bf n}s^{+1}_{\bf n}\rangle=0,
\label{constraint}\end{equation}
which can be written in the form $N^{-1}\sum_{\bf k}C_{\bf k}=1/2$ for
small hole concentrations $x=\langle X_{\bf n}\rangle$. In this equation
and in $C_p$ the value of $C_{\bf k}$ is determined by the equation
\begin{equation}
C_{\bf k}=4J|C_1|(1-\gamma_{\bf k})\omega_{\bf k}^{-1}\coth[
\omega_{\bf k}/(2T)],
\label{ssys}\end{equation}
which follows from Eq.~(\ref{sgf}).

The parameter $\Delta$ describes a gap near the point $(\pi,\pi)$ of the
Brillouin zone in excitation spectrum (\ref{freq}). As follows from our
calculations, the gap opens at $x \approx 0.01$ for $T=0$ and, in accord
with the Mermin-Wagner theorem \cite{mermin}, for any nonzero $T$. The
gap opening indicates the destruction of the long-range
antiferromagnetic order and the establishment of the short-range order
with the correlation length which is determined by the gap magnitude.
Notice that Eq.~(\ref{po}) is close in its form to the polarization
operator obtained in Ref.~\cite{sherman98} for the same model in the
spin-wave approximation. However, the effective interaction constant
$f^2_{\bf k'k}\omega^{-1}_{\bf k}$ differs somewhat from the constant of
the spin-wave approximation.

As can be shown by analogous calculations of the transversal spin
Green's function $D_\perp({\bf k}t)=-i\theta(t)\langle s_{\bf
k}^{-1}(t)s_{\bf k}^{+1}\rangle$,
\begin{equation}
D_\perp({\bf k}t)=2D({\bf k}t),
\label{sphsym}\end{equation}
as it has to be in the paramagnetic state. Thus, the approximations
made do not violate the rotation symmetry of the solution.
Equation~(\ref{sphsym}) was used in the derivation of Eq.~(\ref{ssys}).

\begin{figure}
\vspace{6cm}
\caption{The spin correlations $Z(l)=4|\langle s^z_{(l,0)}s^z_{(0,0)}
\rangle|$ calculated for $T/J=0.5$, 0.75 and 1 in this work (open 
circles) and by the Monte Carlo method in Ref.~\protect\cite{makivic}
(filled circles). In both calculations a 32$\times$32 lattice without
holes was used.}
\label{figi}\end{figure}

\begin{figure}
\vspace{4.7cm}
\caption{The hole spectral function $A({\bf k}\omega)$ for the case of
one hole in a 4$\times$4 lattice, ${\bf k}=(\pi/2,\pi/2)$ and parameters
$J=0.2t$, $\eta=0.1t$. Our calculations for $T=0.02t$ and 
zero-temperature exact-diagonalization data of
Ref.~\protect\cite{dagotto} are given in panels (a) and (b),
respectively.}
\label{figii}\end{figure}

The hole Green's function reads
\begin{equation}
G({\bf k}\omega)=\phi[\omega-\varepsilon_{\bf k}-\mu-
 \Sigma({\bf k}\omega)]^{-1},
\label{hgf}\end{equation}
where $\phi=(1+x)/2$, the unrenormalized hole energy for moderate $x$
\begin{equation}
\varepsilon_{\bf k}\approx\left(1-4\alpha^2C_1^2+4C_1\right)\phi^{-1}t
 \gamma_{\bf k},
\label{urf}\end{equation}
and the imaginary part of the hole self-energy
\begin{eqnarray}
&&{\rm Im}\,\Sigma({\bf k}\omega)=\pi\sum_{\bf k'}h_{\bf kk'}
 \int^\infty_{-\infty}d\omega'[n_B(\omega')+n_F(\omega-\omega')]
 \nonumber\\
&&\qquad\qquad\;\;\times A({\bf k-k'},\omega-\omega')B({\bf k'}\omega'),
\label{se}\end{eqnarray}
with $h_{\bf kk'}=32(N\phi)^{-1}t^2\{(\gamma_{\bf k-k'}+\gamma_{\bf
k})^2+(\gamma_{\bf k-k'}^2-\gamma_{\bf k}^2)[(1+\gamma_{\bf k'})/(1
-\gamma_{\bf k'})]^{1/2}\}$, $n_B(\omega)=[\exp(\omega/T)-1]^{-1}$ and
the spin spectral function $B({\bf k}\omega)=-\pi^{-1}$ ${\rm Im}\,
D({\bf k}\omega)$. It may be noticed that the general structure of
Eq.~(\ref{se}) coincides with the respective equation in the spin-wave
approximation \cite{sherman98} where the interaction constant is close
to $h_{\bf kk'}$. From Eq.~(\ref{ssys}) for an infinite lattice and
small $x$ and $T$ we find $C_1=-0.206734$ and $\alpha=1.70494$ (this
value of $C_1=\langle s^{-1}_{\bf n}s^{+1}_{\bf m}\rangle$ with {\bf n}
and {\bf m} being nearest neighbors is close to the value obtained in
Monte Carlo simulations \cite{makivic}). With these parameters
$\varepsilon_{\bf k}\approx -0.65t\gamma_{\bf k}$. Thus the width of the
unrenormalized band is much smaller than the bandwidth in the absence of
correlations when the dispersion is described by $4t\gamma_{\bf k}$.
This difference stems from the antiferromagnetic alignment of spins
which leads to spin flipping accompanying the hole motion.

To check the validity of the approximations made we carried out a number
of calculations which allowed comparison with exact diagonalisation and
Monte Carlo data. In Fig.~\ref{figi} results of our calculations of the
spin correlations are compared with Monte Carlo data of
Ref.~\cite{makivic}. In Fig.~\ref{figii} the hole spectral function
$A({\bf k}\omega)$ obtained from our calculations and by exact
diagonalization in Ref.~\cite{dagotto} are compared. In both
calculations the artificial broadening $\eta=0.1t$ was introduced to
transform $\delta$-functions into Lorentzians. As is seen from these
figures, our calculations are in good agreement with the numerical
experiments.

\begin{figure}
\vspace{7.5cm}
\caption{The hole spectral function along the $(0,0)-(\pi,0)$ direction
of the Brillouin zone. The calculations were carried out on a
20$\times$20 lattice for the parameters $t=0.5$~eV, $J=0.1$~eV,
$x\approx 0.1$ and $T=0$. The Fermi energy is taken as zero of energy.
In contrast to Fig.~\protect\ref{figii} the electron picture is used in
this figure.}
\label{figiii}\end{figure}

The mentioned similarity of Eqs.~(\ref{sgf}), (\ref{po}), (\ref{hgf})
and (\ref{se}) to the respective equations of the spin-wave
approximation \cite{sherman98} at moderate doping leads to close
resemblance in shape of the spectral functions obtained in these two
approaches. As an example of the obtained results, the hole spectral
function is shown in Fig.~\ref{figiii} for values of $t$ and $J$ which
correspond to parameters of cuprate perovskites. As seen from the
figure, for moderate doping the hole spectrum contains two dispersive
features for wave vectors near the boundary of the magnetic Brillouin
zone. One of these features -- a narrow intensive peak slightly below
(in the electron picture) the Fermi level $\omega=0$ -- is connected
with the so-called spin-polaron band. In a lightly doped crystal the
width of this band is of the order of the exchange constant $J$ which is
much smaller than the hopping constant $t$ for parameters of cuprate
perovskites. The spin-polaron bandwidth is characterized by the
parameter of magnetic excitations because on the antiferromagnetic
background the hole movement is accompanied by spin flipping. The
short-range antiferromagnetic ordering is retained in moderately doped
crystals. In these conditions a part of the spin-polaron band is
preserved near the boundary of the magnetic Brillouin zone. The second
dispersive feature is a broad maximum the dispersion of which is
characterized by the second energy parameter of the model $t$. The shape
of this dispersion reproduces with some distortion the shape of the
two-dimensional nearest-neighbor band. For wave vectors near the
$\Gamma$ point two broad maxima, one below and one above the Fermi
level, are seen in Fig.~\ref{figiii}. Both of them belong to the
dispersion with the characteristic energy $t$.

\begin{figure}
\vspace{10cm}
\caption{(a) The Fermi surface for the underdoped case. (b) The hole
spectral functions for wave vectors on the dashed curve of the Fermi
surface in panel (a). Curves from top to bottom correspond to ${\bf
k}=(0.2\pi,\pi)$, $(0.2\pi,0.9\pi)$, $(0.2\pi,0.8\pi)$, $(0.3\pi,0.7
\pi)$, $(0.4\pi,0.6\pi)$ and $(0.5\pi,$ $0.5\pi)$, respectively.
$T=116$~K, $x=0.121$, $t=0.5$~eV, and $J=0.1$~eV. (c) The energy gap
between the spin-polaron peak and the Fermi level along the dashed curve
of the Fermi surface for $x=0.121$ (squares) and $x=0.171$ (diamonds) at
$T=116$~K . The position of the leading edge of the photoemission
spectrum measured \protect\cite{ding} in underdoped
Bi$_2$Sr$_2$CaCu$_2$O$_{8+\delta}$ along the similar Fermi surface is
indicated by filled circles.
} 
\label{figiv}\end{figure}

The Fermi surface derived from the spectral functions in the underdoped
case is shown in Fig.~\ref{figiv}a. The solid segments along the
boundary of the magnetic Brillouin zone are formed by points where the
spin-polaron band touches the Fermi level (notice that this band does
not cross it). The Fermi level is crossed by the broader dispersive
feature along the dashed curves. The spectral functions calculated for
wave vectors on these curves are shown in Fig.~\ref{figiv}b. On moving
from $(\pi/2,\pi/2)$ to $(\pi/5,\pi)$ on the boundary of the Brillouin
zone the spin-polaron peak recedes from the Fermi level. The situation
looks like no Fermi level crossing occurs near $(\pi/5,\pi)$ and a gap
opens between the energy band and the Fermi level. Only on another scale
(compare Fig.~\ref{figii}) one can see that in this region the Fermi
level crossing does exist. It is not seen in Fig.~\ref{figiv}b because
the maximum which crosses the Fermi level is completely lost at the foot
of the more intensive spin-polaron peak and this peak is positioned
somewhat below the Fermi level. Analogous behavior of the photoemission
leading edge is observed in cuprate perovskites and is known as the
photoemission pseudogap \cite{ding,shen}. Notice that the shape of the
photoemission spectrum is determined by the spectral function and the
correlation between their behaviors is expected. In Fig.~\ref{figiv}c
the positions of the spin-polaron peak in the spectral functions and the
leading edge of the photoemission spectrum measured \cite{ding} in
underdoped Bi$_2$Sr$_2$CaCu$_2$O$_{8+\delta}$ are compared for wave
vectors on the Fermi surface. The magnitude of the pseudogap calculated
for $x=0.121$ and its symmetry are close to those observed
experimentally. Moreover, at $x=0.171$ which is close to the optimal
doping concentration the calculated pseudogap disappears. This is also
in agreement with experiment.

\begin{figure}
\vspace{8cm}
\caption{(a) The normal state photoemission spectrum of the optimally
doped Bi$_2$Sr$_{2-x}$La$_x$CuO$_{6+\delta}$ ($T_c=29$~K) at $T=45$~K
and ${\bf k}=(0.62\pi,0)$ \protect\cite{janowitz}. (b) The photoemission
spectrum calculated from the spectral function of the $t$-$J$ model for
$t=0.5$~eV, $J=0.1$~eV, $x\approx 0.1$, $T=45$~K and ${\bf
k}=(0.6\pi,0)$. (c) The dispersions of the two features in the
photoemission spectrum of Bi$_2$Sr$_{2-x}$La$_x$CuO$_{6+\delta}$
\protect\cite{janowitz}. (d) The dispersions of the maxima in the
spectral function in Fig.~\protect\ref{figiii}.} 
\label{figv}\end{figure}

Two distinct emissions were observed recently in the normal state
photoemission of the optimally doped
Bi$_2$Sr$_{2-x}$La$_x$CuO$_{6+\delta}$ in the vicinity of the point
$(\pi,0)$ \cite{janowitz}. This crystal is a single CuO$_2$ layer
material with sufficiently decoupled layers. Therefore the two emissions
cannot be ascribed to the bilayer splitting. As mentioned, the hole
spectral function of the two-dimensional $t$-$J$ model has also two
components and it is reasonable to compare them with experimental data.
The photoemission spectrum in Fig.~\ref{figv}(b) was obtained from the
spectral function by convoluting it with a Gaussian to model
experimental resolution and cutting off with the Fermi distribution. In
accord with experimental conditions of Ref.~\cite{janowitz}, the
Gaussian width 20~meV and the temperature 45~K were selected. In spite
of some differences in relative intensities and widths of the spectral
features, the calculated spectrum has the same structure as the
experimental spectrum shown in panel (a). As in experiment, the maximum
with a larger binding energy has a considerably larger width and
stronger dispersion in comparison with the maximum with a lower binding
energy. The experimental and calculated dispersions of the two spectral
features are compared in panels (c) and (d). As is seen from the figure,
they are similar and the binding energies are of the same order of
magnitude. Recently two dispersive features were also resolved in the
normal-state photoemission of the underdoped and optimally doped
Bi$_2$Sr$_2$CaCu$_2$O$_{8+\delta}$ with two CuO$_2$ planes in the unit
cell \cite{Chuang}. This splitting was connected with a coupling between
the adjacent CuO$_2$ planes. Since energy distribution curves and values
of the band splitting in that work are similar to those observed in the
single CuO$_2$ plane material Bi$_2$Sr$_{2-x}$La$_x$CuO$_{6+\delta}$,
another possible mechanism of the band splitting in
Bi$_2$Sr$_2$CaCu$_2$O$_{8+\delta}$ -- the above-discussed peculiarity of
the spectrum of the two-dimensional strongly correlated system -- has to
be also taken into account. It counts in favour of this latter mechanism
that the Fermi surface obtained in Ref.~\cite{Chuang} (see Fig.~1g
there) is very close to the Fermi surface in Fig.~\ref{figiv}a.

In summary, the $t$-$J$ model of the CuO$_2$ plane was investigated with
the method which retains the rotation symmetry of the paramagnetic state
and accounts properly for the kinematic interaction. The observed
unusual properties of cuprates -- the photoemission pseudogap and the
band splitting -- were shown to be inherent in the two-dimensional
strongly correlated electron system.

\acknowledgements
This work was partially supported by the ESF grant No.~4022 and by the
WTZ grant (Project EST-003-98) of the BMBF.

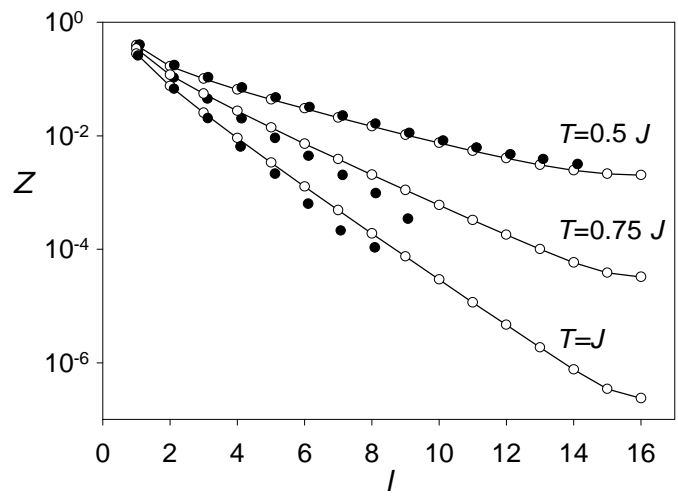
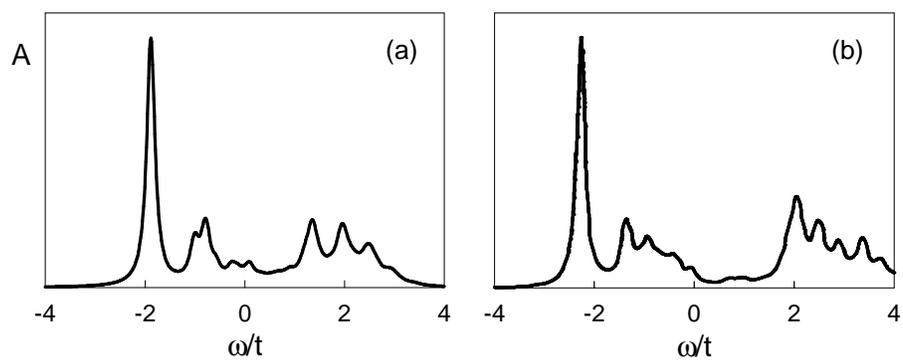
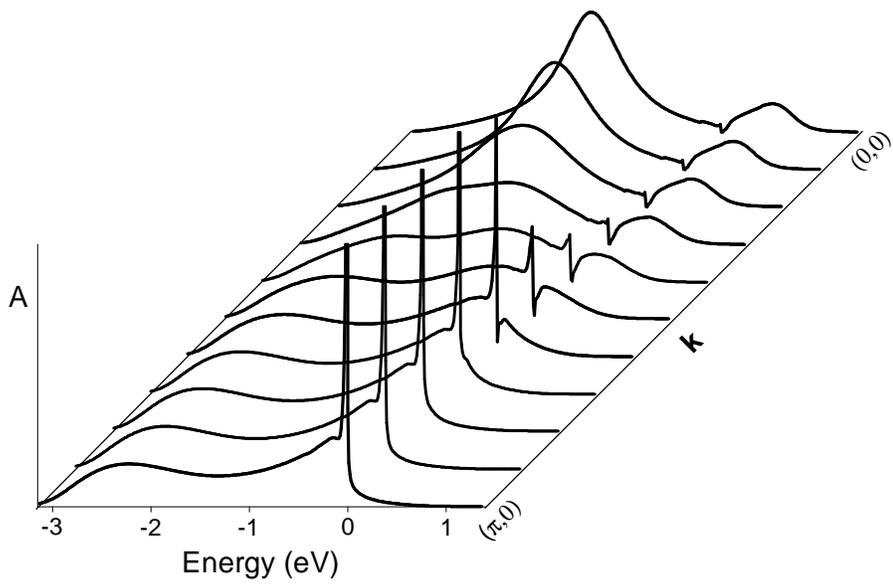
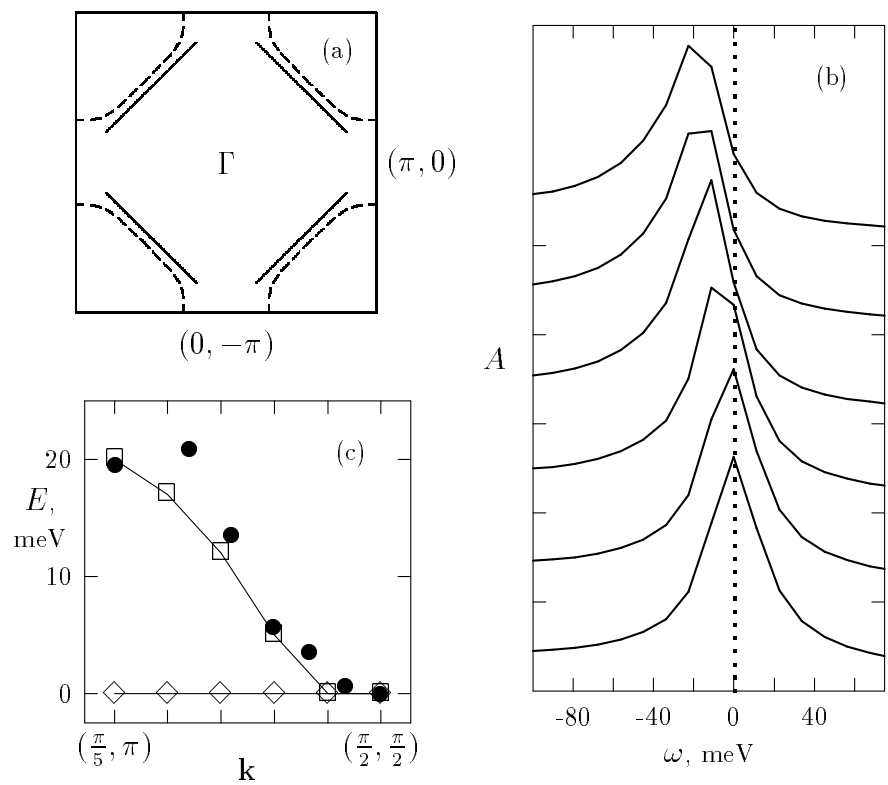
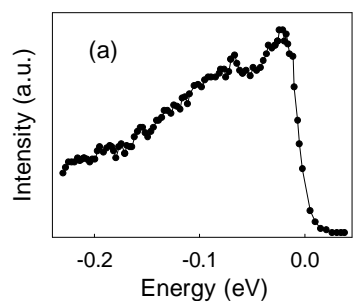
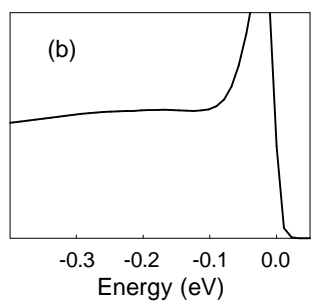
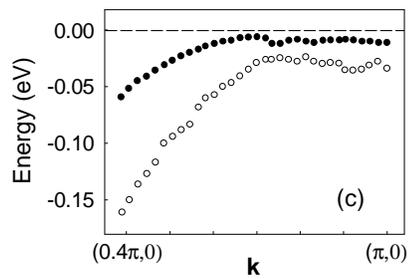
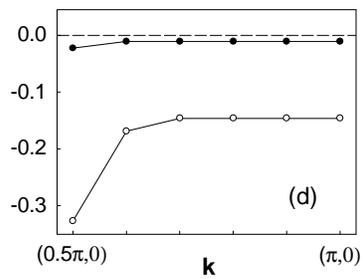
\end{document}